\documentclass[prd,preprintnumbers,twocolumn,eqsecnum,floatfix,a4paper,nofootinbib,superscriptaddress]{revtex4-1}

\usepackage{color}
\usepackage{amsmath,amssymb,graphicx}
\usepackage{bm,lipsum}
\usepackage{times}
\usepackage{microtype}
\usepackage[utf8]{inputenc}
\usepackage{booktabs}
\usepackage{subfigure}
\usepackage[normalem]{ulem}
\usepackage[varg]{txfonts}
\usepackage{bm,graphics,graphicx,epsfig,xcolor,ulem}
\usepackage[colorlinks, pdfborder={0 0 0}]{hyperref}
\usepackage{multirow}
\definecolor{LinkColor}{rgb}{0.75, 0, 0}
\definecolor{CiteColor}{rgb}{0, 0.5, 0.5}
\definecolor{UrlColor}{rgb}{0, 0, 0.75}
\hypersetup{linkcolor=LinkColor}
\hypersetup{citecolor=CiteColor}
\hypersetup{urlcolor=UrlColor}
\allowdisplaybreaks
\begin{document}

\title{Tests of general relativity using multiband
observations of intermediate mass binary black hole mergers}
\author{Sayantani Datta} \email{sdatta94@cmi.ac.in} 
\affiliation{Chennai Mathematical Institute, Siruseri, 603103, India}
\author{Anuradha Gupta}  \email{agupta1@olemiss.edu}
\affiliation{Department of Physics and Astronomy, The University of Mississippi, Oxford, Mississippi 38677, USA}
\affiliation{Institute for Gravitation and the Cosmos, Department of Physics, Penn State University, University Park, Pennsylvania 16802, USA}
\author{Shilpa Kastha} \email{shilpa.kastha@aei.mpg.de}
\affiliation{Max Planck Institute for Gravitational Physics (Albert Einstein Institute), Callinstra\ss e 38, D--30167 Hannover, Germany \\ Leibniz Universit\" at Hannover, D--30167 Hannover, Germany \\}
\author{K. G. Arun}  \email{kgarun@cmi.ac.in} 
\affiliation{Chennai Mathematical Institute, Siruseri, 603103, India}
\affiliation{Institute for Gravitation and the Cosmos, Department of Physics, Penn State University, University Park, Pennsylvania 16802, USA}
\author{B. S. Sathyaprakash}  \email{bss25@psu.edu} 
\affiliation{Institute for Gravitation and the Cosmos, Department of Physics, Penn State University, University Park, Pennsylvania 16802, USA}
\affiliation{Department of Astronomy and Astrophysics, Penn State University, University Park, Pennsylvania 16802, USA}
\affiliation{School of Physics and Astronomy, Cardiff University, Cardiff, CF24 3AA, United Kingdom}

\date{\today}

\begin{abstract}
Observation of gravitational waves (GWs) in two different frequency bands is referred to as {\it multiband GW astronomy}.  With the planned Laser Interferometric Space Antenna (LISA) operating in the $10^{-4}-0.1$ Hz range, and third generation (3G) ground-based detectors such as the Cosmic Explorer (CE) and Einstein Telescope (ET), operating in the $1$--$10^4$ Hz range, multiband GW astronomy could be a reality in the coming decades. In this paper we present the potential of multiband observations of intermediate mass binary black holes (IMBBHs) of component masses ${\sim}10^2$--$10^3M_{\odot}$ to test general relativity (GR). We show that mutiband observations of IMBBHs would permit multiparameter tests of GR---tests where more than one post-Newtonian (PN) coefficient is simultaneously measured---yielding more rigorous constraints on possible modifications to GR. We also find that the improvement due to multibanding can often be much larger than the best of the bounds from either of the two observatories. The origin of this result, as we shall demonstrate, can be traced to the lifting of degeneracies among the various parameters when the information from LISA and 3G is taken together. A binary of redshifted total mass of $200 M_{\odot}$ gives the best bounds. Such a system at 1 Gpc and mass ratio $m_1/m_2=2$ would bound the deviations on all the PN coefficients to below 10\% when they are measured individually, and additionally place simultaneous bounds on the first seven PN coefficients to below  50\%.   
\end{abstract} 
\maketitle

\section{Introduction} 

Einstein's general relativity (GR) has been subjected to a plethora of tests performed in the laboratory as well as those using
astrophysical observations~\cite{WillLR05}. The theory has so far been consistent with each of these tests
(see Refs.~\cite{YunesSiemens2013,GairLivRev, SathyaSchutzLivRev09,Berti:2015itd,APrev13} for an overview of various 
astrophysical tests of GR). The first observation of gravitational waves (GWs) from the binary black hole (BBH) merger GW150914~\cite{Discovery}
 and several others~\cite{GW151226,GW170104,GW170608,GW170814,GWTC1,Venumadhav:2019lyq,Zackay:2019btq,Abbott:2020niy}  during the first, second, 
and the first half of the third observing runs, have permitted tests of GR in a regime of 
strong gravity and high curvature which had previously been elusive~\cite{GWTC-TGR,Abbott:2020jks}. 
 The binary neutron star merger GW170817~\cite{GW170817} further facilitated tests of strong-field gravity for non-vacuum
spacetimes~\cite{BNSTGR}. The observation of electromagnetic emission associated with this event helped in deriving stringent
 constraints on the speed of the GWs~\cite{GW-GRB170817}. All these tests have placed tighter constraints on possible deviations
 from GR \cite{YYP2016} while ruling out modified theories of gravity invoked to explain dark energy \cite{Baker:2017hug,Creminelli:2017sry,Sakstein:2017xjx,Ezquiaga:2017ekz,Boran:2017rdn}.  
The recent detection of a BBH merger, GW190521 \cite{GW190521}, with total mass $\sim150M_{\odot}$ has opened up new possibilities for understanding the formation mechanisms of BBHs as well as tests of GR.

\emph{Parametrized tests of GR}~\cite{AIQS06a, AIQS06b, MAIS10, YunesPretorius09, LiEtal2011, TIGER}, 
are among the pioneering tests of the theory performed with GW data. These tests make the best use of the structure of the GW phase evolution from the post-Newtonian (PN) approximation to GR~\cite{Bliving}. 
In the PN approximation, the phase evolution of the GW signal 
 can be expanded as a power series in $v$ and $\log v$, where $v$ denotes the velocity parameter describing the orbital motion of the binary.
The different PN orders (corresponding to different powers of $v$) capture the diverse physics and various nonlinear effects underlying the compact
binary dynamics. Hence, looking for deviations in the PN coefficients is equivalent to constraining the different physics that goes into
them~\cite{BSat94,BSat95}. In this framework, deviations from GR are parametrized via deformation in the phasing formula at different PN orders~\cite{LiEtal2011,TIGER},  
whose values are put to test using the GW data. As these deformation parameters take the value zero in GR, this null test is devised to derive constraints on them at a fixed credible level.

The parametrized tests of GR branch out into several subclasses depending on the number of PN deformation parameters that are simultaneously
estimated from the data. Ideally, one aims to constrain all or several of the PN deformation parameters simultaneously using the GW data~\cite{AIQS06a}. Tests that do this will be referred to as {\it multiparameter tests} in this paper. 
One may wish to further classify these multiparameter tests into two classes, depending on whether the block of PN parameters that are tested starts from the lowest PN order (in the ascending order) or from the highest
PN order (in the descending order). The former would make sense in terms of verifying the predictions of GR at different PN orders with increasing levels of complexity in the nonlinear interactions.
 The latter perspective, starting from the highest PN order and proceeding in decreasing order, 
would be expected from modified theories such as an effective field theory where modifications to GR would start at a particular PN order and all
orders above that \cite{Endlich:2017tqa,Sennett:2019bpc}. There could be other possible combinations of the PN deformation parameters  that may be tested simultaneously, 
but we consider only these two classes of the multiparameter tests in this paper as they are the most general ones.
These classes of tests, though more rigorous, yield weaker bounds, compared to single-parameter tests, due to the large correlations of the deformation parameters among themselves as well as with the intrinsic parameters
 of the binary, such as its masses and spins~\cite{AIQS06a, Carl:multiparam}.

Hence, one considers a somewhat less rigorous set of tests where only one of the many PN deformation parameters is chosen at a time as a test parameter~\cite{AIQS06b, LiEtal2011,TIGER}. This is less rigorous in the sense that a modification to the phasing formula from a non-GR theory is likely to occur at more than one PN order. This aspect is not accounted for in the formulation of the single-parameter tests. This drawback is partially compensated by performing a set of tests while varying the PN deformation parameter systematically from 0PN till 3.5PN one-by-one (see \cite{MAIS10} for a detailed discussion). Hence, one or more of these tests would potentially detect a deviation if the underlying theory of gravity is not GR, though a deviation seen at a particular PN order in this test does not necessarily mean the breakdown of GR occurs only at  that particular order. Given the sensitivity of the current generation of GW detectors, this is the method presently being  employed in the analysis of the LIGO and Virgo data and will be referred to as {\it single-parameter tests} in this paper. The current constraints from the single-parameter tests, at a 90\% credible level, from the BBHs observed so far from LIGO and Virgo detectors are reported in \cite{GWTC-TGR,Abbott:2020jks}. However, with the next-generation ground-based and space-based detectors, the sensitivity would reach levels where the
multiparameter tests would be possible~\cite{Gupta:2020lxa}. 

Several studies have quantified the projected bounds on these PN parameters using third generation (3G) ground-based GW experiments such as the Einstein Telescope (ET) \cite{Punturo:2010zz} and Cosmic Explorer (CE) \cite{2019arXiv190704833R} as well as the space-based Laser Interferometer Space Antenna (LISA) \cite{LISA2017} (see, e.g., \cite{AIQS06a,AIQS06b,MAIS10,Chamberlain:2017fjl}). The ground-based 3G detectors are sensitive to  stellar mass BBHs (up to ${\sim} 100 M_{\odot}$) and intermediate mass BBHs (IMBBHs) (${\sim} 10^2-10^3 M_{\odot}$) in the frequency range
${\sim} 1-10^4$ Hz~\cite{CEDwyer,3Gsens}, while the space-based LISA mission is most sensitive to supermassive black hole mergers
 (${\sim} 10^{5-7} M_{\odot}$) in the $10^{-4}-0.1$ Hz band~\cite{Babak2017}.  While CE/ET will have the advantage of around 10--40 fold improvement in strain sensitivity compared to the present generation detectors such as advanced LIGO and advanced Virgo, LISA will open up the possibility of
using mHz band for GW astronomy. But neither of them may be able to set stringent enough bounds on the deformation parameters to rule in or rule out viable modified theories of gravity~\cite{Gupta:2020lxa}. This is because the intrinsic degeneracies of the deformation parameters with masses and spins prevent making precise measurements of these parameters using either ground-based or space-based experiments alone
~\cite{MAIS10}. 

In the past few years, an alternative strategy to combine these two classes of observations has been proposed as a new tool to probe the
strong-field dynamics~\cite{Nair:2015bga,Vitale:2016rfr,Barausse_2016,Carson_2019,Gnocchi_2019,Grimm:2020ivq,Toubiana:2020vtf} of BBHs. 
This is often referred to as {\emph {multiband GW astronomy}} \cite{Sesana:2016ljz} where, using a class of sources visible in both LISA and 3G, 
one combines the low frequency content (the early dynamics of compact binaries) in the LISA band and the high-frequency content
(carrying an imprint of the late time dynamics of compact binaries) in the 3G band to obtain bounds on departures from GR. Various studies on multiband parameter estimation mostly used stellar-mass BBHs like GW150914, 
which will have a signal-to-noise ratio (SNR) of order unity in the LISA band, but a ratio of several hundreds to thousands in the 3G band. Even then, in these studies, joint observation has been argued to be able to provide bounds several orders of magnitude better than those from individual observations~\cite{Barausse_2016,Carson_2019,Gnocchi_2019, Grimm:2020ivq,Toubiana:2020vtf}. This huge improvement has been broadly attributed to the combination of the low-frequency sensitivity of LISA with the high-frequency response of the 3G detectors. In Ref.~\cite{Toubiana:2020vtf}, these generic features have been confirmed, for the first time, with a treatment of the problem within the Bayesian
inference framework.
\subsection{Multiband tests of GR using IMBBHs}
In this work, we take this paradigm forward by carrying out an extensive study of the effect of multiband observations of IMBBHs, as opposed to stellar mass BBHs,  using LISA and CE/ET,  
with total source-frame masses of the binaries varying between $100-550 M_{\odot}$\footnote{{\label{550}} This choice of masses is made to ensure multiband visibility of the GW inspiral from the sources and is not from any astrophysical consideration.}. Astrophysically, IMBHs can have masses as high as $10^{4}M_{\odot}$~\cite{Miller:2003sc} and would be excellent sources of GWs for these detectors. Therefore, the detecting and understanding the formation of IMBBHs are among the top science priorities of present and future astronomical telescopes~\cite{Bellovary:2019nib,Wrobel:2019luy}. 
The multiband detectability of IMBBHs is discussed in detail in Ref.~\cite{Jani:2019ffg}. Further, possible implications for the multibanding
of IMBBHs for parameter estimation and tests of GR are highlighted in \cite{Cutler:2019krq}. Among the LIGO/Virgo detections so far \cite{Abbott:2020niy}, GW190521, at a redshift of $0.82^{+0.28}_{-0.34}$, is the most massive BBH, with a total source-frame-mass of $150^{+29}_{-17} M_{\odot}$ and a total median redshifted-mass of $\sim 270 M_{\odot}$. Though BBHs in this mass range are ideal for multibanding~\cite{Ezquiaga:2020tns}, due 
 to the relatively high redshift of the source, GW190521 would have an SNR of $\sim2-5$ in the LISA band, making its detection unlikely. In this paper we present a detailed study of the implications
of multiband observations of IMBBHs by LISA and CE/ET detectors in terms of tests of GR.

Unlike the stellar-mass BBHs, the IMBBHs will have SNRs of the order of tens in the LISA band (as opposed to an order of unity for
stellar-mass BBHs) while in the CE/ET band they still have SNRs of the order of hundreds to thousands. This significantly helps the process of multibanding thereby providing us precise measurements of the PN deformation parameters. Further, our detailed study reveals that the dramatic improvements from multibanding are due to large scale cancellations 
 of correlations among different parameters in the problem, due to the mutual complementarity of the two experiments or frequency bands. 
We observe that at an intermediate mass BBH with a total redshifted-mass of $200 M_{\odot}$ is a sweet spot for multiparameter tests with multiband observations.
Considering such a system of mass ratio 2  and at 1 Gpc, we find that single-parameter tests can constrain the first three PN deformation parameters to accuracy $\sim 0.1\%$ and the rest to below $10\%$. Multiparameter tests up to a 7-parameter case can be performed with the above system, where the first two PN deformation parameters are bounded to below $0.5\%$ and the rest to below $50\%$ (with low spins).
 
The remainder of the paper is organized as follows: In Sec.~\ref{sec:multibandTGR} we discuss the basic concepts in combining the
information from the two frequency bands (LISA and 3G), specific to the case of parametrized tests of GR. The results for the 
single-parameter tests with explanations of the trends seen are presented in Sec. \ref{sec:singleparam}. Multiparameter tests are discussed in Sec.~\ref{sec:multiparam}. 
Lastly, some of the caveats of the analysis are listed in Sec.~\ref{sec:caveats} and our conclusions are provided in Sec.~\ref{sec:conclusion}.
%%
%%%
\section{Tests of GR using multiband GW
observations}\label{sec:multibandTGR}
\subsection{Parametrized tests of GR using IMRPhenomD waveforms}\label{sec:IMRPD}
 The breakthrough in numerical relativity~\cite{Pretorius05,Pretorius07Review} has enabled us to construct analytical or semianalytical waveforms which account for the
  inspiral (the early phase of binary evolution), merger (the late stages of the binary evolution as the two objects coalesce) and ringdown (the post-merger phase of the remnant black hole) phases~\cite{Ajith09, Husa2016, Khan2016, Khan:2018fmp, Khan:2019kot, BuonD98, BuonEOB07}. An important subclass of them, referred to as inspiral-merger-ringdown phenomenological 
  waveforms or IMRPhenom, is constructed starting with an ansatz about the structure of the frequency domain gravitational waveforms, which contain several free parameters that are fixed by matching with numerical relativity simulations for various mass-ratios and spins. Here we use the IMRPhenomD waveform model~\cite{Khan2016} of the IMRPhenom family, which assumes the spins of the binary constituents to be aligned or anti-aligned with respect to the orbital angular momentum vector and hence the binary is non-precessing. The amplitude of IMRPhenomD accounts for only the leading 
  quadrupolar ($l = 2, |m| = 2$) mode. Schematically a frequency domain waveform would read
\begin{equation}
{\tilde h}(f)={\cal A}(f) \, e^{i\Phi (f)},
\end{equation}
where ${\cal A}(f)$ and $\Phi (f)$ denote the amplitude and phase of the
gravitational waveform. The phase, in the inspiral regime, admits an
expansion of the form~\cite{Bliving,BDI95,BDIWW95,BFIJ02,BDEI04,ABFO08,MKAF16}
\begin{equation}\label{schematic-PN-phasing}
\Phi(f)=2\pi f\,t_c-\phi_c+\frac{3}{128\,\eta\,v^5}\left[ \sum_{k=0}^K
\phi_k\,
v^k+\sum_{kl=0}^K\phi_{kl}\,v^{kl}\ln v\right],
\end{equation}
where $t_c,\phi_c$ are kinematical parameters related to the time and phase of the arrival of the signal at the detector and $v \equiv (\pi Mf)^{1/3}$ is the  PN expansion parameter in terms of which the amplitude and phase are expressed. Furthermore, $\eta \equiv m_1 m_2/M^2$ is the symmetric mass ratio where $M \equiv m_1 + m_2$ is the total mass of the binary. Note that these masses are the detector
frame (or redshifted) masses after accounting for the redshift of the source and related to the source-frame masses $M_{\rm source}$ by $M=M_{\rm source}(1+z)$. The PN coefficients in the phasing formula, deformations in which we are interested, are denoted by $\phi_k$ and $\phi_{kl}$.

For the present analysis, we use the amplitude ${\cal A}(f)$ and phase $\Phi (f)$ of the IMRPhenomD waveforms, which by construction agree with the predictions of 
PN theory in the inspiral part of the waveform. The inspiral part of the GW phasing is described by the  PN
phasing formula, in Eq.~(\ref{schematic-PN-phasing}), correct to 3.5PN order (${\cal O} (v^7)$). All of the PN coefficient $\phi_k$ and $\phi_{kl}$ are functions of the various combinations of the intrinsic parameters of the system, such as the total mass $M$, 
the symmetric mass ratio $\eta$ and the dimensionless spins $\chi_{1,2}$ of the binary components. 
Any modification to these PN coefficients  would potentially arise from modifications to GR, an assumption which forms the basic premise for the parametrized tests. We take the ansatz for PN expansion present in IMRPhenomD and introduce free parameters at every PN order to model non-GR modifications, rewriting the PN coefficients as,
\begin{eqnarray} \label{eq:dev}
\phi_k  \rightarrow  \phi_k( 1+\delta{\hat \phi}_k ),\\
\phi_{kl}  \rightarrow  \phi_{kl}( 1+\delta{\hat \phi}_{kl} ),
\end{eqnarray}
where $\delta\hat \phi_k$ ($k = 0, 2, 3, 4, 6, 7$) and  $\delta\hat \phi_{kl}$ ($kl = 5l, 6l$) are the fractional non-GR deformation parameters to the respective PN orders denoted by the indices $k$ and $kl$.  More specifically, $5l$ and $6l$ represent the deformations to the logarithmic terms at 2.5PN and
3PN, respectively. The non-logarithmic part of the 2.5PN order coefficient does not have any frequency dependence and can be absorbed by redefining $\phi_c$. This is why we do not consider $\delta\hat \phi_{5}$ as a separate parameter. Since the gravitational waveform is intrinsically parametrized by $M,$ $\eta$ and $\chi_{1,2}$ as discussed above, our parameter space is 15 dimensional, consisting of seven GR parameters (including the  luminosity distance $D_L$ of the source) and eight non-GR deformation parameters, $\{\delta{\hat \phi}_k\}$ and $\{\delta{\hat \phi}_{kl}\}$:
\begin{equation}
{\mathbf {\theta}}^{\:a}=\{{{\rm ln}\: D_L},\:t_c,\: \phi_c,\:{\rm ln}\:M_c,\:\eta,\:\chi_1,\:\chi_2, \{\delta{\hat
\phi}_k\}, \{\delta{\hat\phi}_{kl}\} \},\label{eq:params}
\end{equation}
where, we find it convenient to use the {\it chirp mass} $M_c \equiv \eta^{3/5} M$ instead of total mass $M$ as one of the mass parameters. We carry out parameter estimation of the GW signal described by these parameters  using the projected sensitivities of  LISA and 3G detectors, the details of which are discussed in the next sections.
\subsection{Detector configurations}\label{sec:detectors}
\subsubsection{Laser Interferometric Space Antenna}
Following the huge success of the LISA-Pathfinder~\cite{LPF}, which has set a benchmark for a millihertz GW experiment in space, the European Space Agency has selected LISA for its L3 mission \cite{Babak2017}. The proposed 
mission, to be launched in 2034, has an equilateral triangular constellation of three spacecraft separated by $2.5\times10^6$ km, connected by six laser links. The constellation would orbit the Sun, trailing behind Earth's orbit at an inclination of 20$^{\circ}$ with respect to the ecliptic. The orbital motion of the spacecraft around the Sun is important for source localization and luminosity  distance estimation~\cite{Cutler98}.

For our purposes, we ignore the orbital motion of LISA. This is justified as our aim is to study the error on the intrinsic parameters of the binary, 
including the PN deformation parameters, which are more or less uncorrelated to the extrinsic parameters (such as luminosity distance, source position and orientation) and hence have minimal effect on our estimates~\cite{BBW05a,ALISA06,Toubiana:2020vtf}. The noise power spectral density (PSD) for LISA consists of the instrumental noise and the confusion noise due to Galactic white dwarf binaries that limit LISA sensitivity in the lower-frequency regime. The instrumental noise that we employ can be found in Eq.~(1) of \cite{Babak2017} and the galactic confusion noise component, corresponding to 4 years of observation time can be found in ~\cite{Mangiagli:2020rwz}.
The LISA noise PSD given in \cite{Babak2017} is averaged over the sky and polarization angles and takes into consideration its triangular shape~\cite{Cutler98,BBW05a,ALISA06,Cornish:2018dyw}. 
We also divide the total noise PSD by 2, taking into account the summation over two independent low-frequency channels. However, to account for averaging over the inclination angle which specifies the orientation of the source with respect to the detector, it will involve an additional prefactor of $\sqrt{4/5}$ multiplying the amplitude of  the IMRPhenomD waveform model~\cite{Cornish:2018dyw}.

The choice of the low frequency cut-off for IMBBHs in the LISA band will depend on the duration over which the signal would
last in the LISA band. If an IMBBH with chirp mass $M_c$ is observed for a duration $T_{\rm obs}$ by LISA, the low frequency cut-off may be chosen as
 \begin{equation}
f_{\rm lower}^{\rm  LISA} = {\rm Max} \left[10^{-4}, 4.149\times 10^{-5}\left(  \frac
{M_c}{10^6 M_{\odot}}\right)^{-5/8} T_{\rm obs}^{-3/8}\right].
\label{flowerLISA}
\end{equation}
The Max argument above ensures that the low-frequency cut-off is not lower than the nominal low frequency cut-off of the LISA instrument. 
In our analysis we take $T_{\rm obs}$ to be 4 years. The upper frequency cut-off is chosen to be 0.1 Hz which is equal to the upper frequency
cut-off the LISA instrument.
\subsubsection{Cosmic Explorer and Einstein Telescope}\label{sec:CE-ET}
We consider Cosmic Explorer and Einstein Telescope as two prototypical detectors representing the sensitivities that will be achieved by the third generation ground-based detectors. 
Cosmic Explorer is a third generation ground-based GW detector proposed in the United States~\cite{2019arXiv190704833R}. It is conceived to be a 40 km L-shaped detector whose science goals are reviewed in \cite{2019arXiv190704833R}. 
We use a noise PSD for CE given in \cite{Kastha2018}.
A similar observing facility is also envisaged in Europe called the Einstein Telescope~\cite{ET-D}. It is a triangular shaped detector
with 10 km arms and is effectively three V-shaped detectors with an opening angle of $60^\circ$. 
Both CE and ET noise sensitivities are limited by gravity gradient noise in the low frequency regime. The lower cut-off frequencies for the ET and CE are chosen to be 1 Hz and 5 Hz, respectively. 
The upper cut-off frequency is chosen such that the characteristic amplitude ($2 {\sqrt f} |\tilde h(f)|$) 
of the GW signal is lower than that of CE/ET noise by $10\%$ at maximum. Though theoretically the upper frequency limit is infinity,
contributions from the high frequency part of the waveform that contribute negligibly are ignored. This choice of the cut-off helps in significantly 
improving the accuracy of the numerical analysis (the SNR computation in the next subsection uses a different upper cut-off for reasons explained there). Similar to the case of LISA, we aim to study the parameter estimation problem using the ET and CE from the standpoint of a test of GR. 
As we use single detector configurations, we multiply the amplitude with a prefactor 2/5 to account for the averaging over the antenna pattern functions~\cite{DIS00,BBW05a,ALISA06}.

\begin{figure}[htp]
\centering
\includegraphics[width=0.52\textwidth]{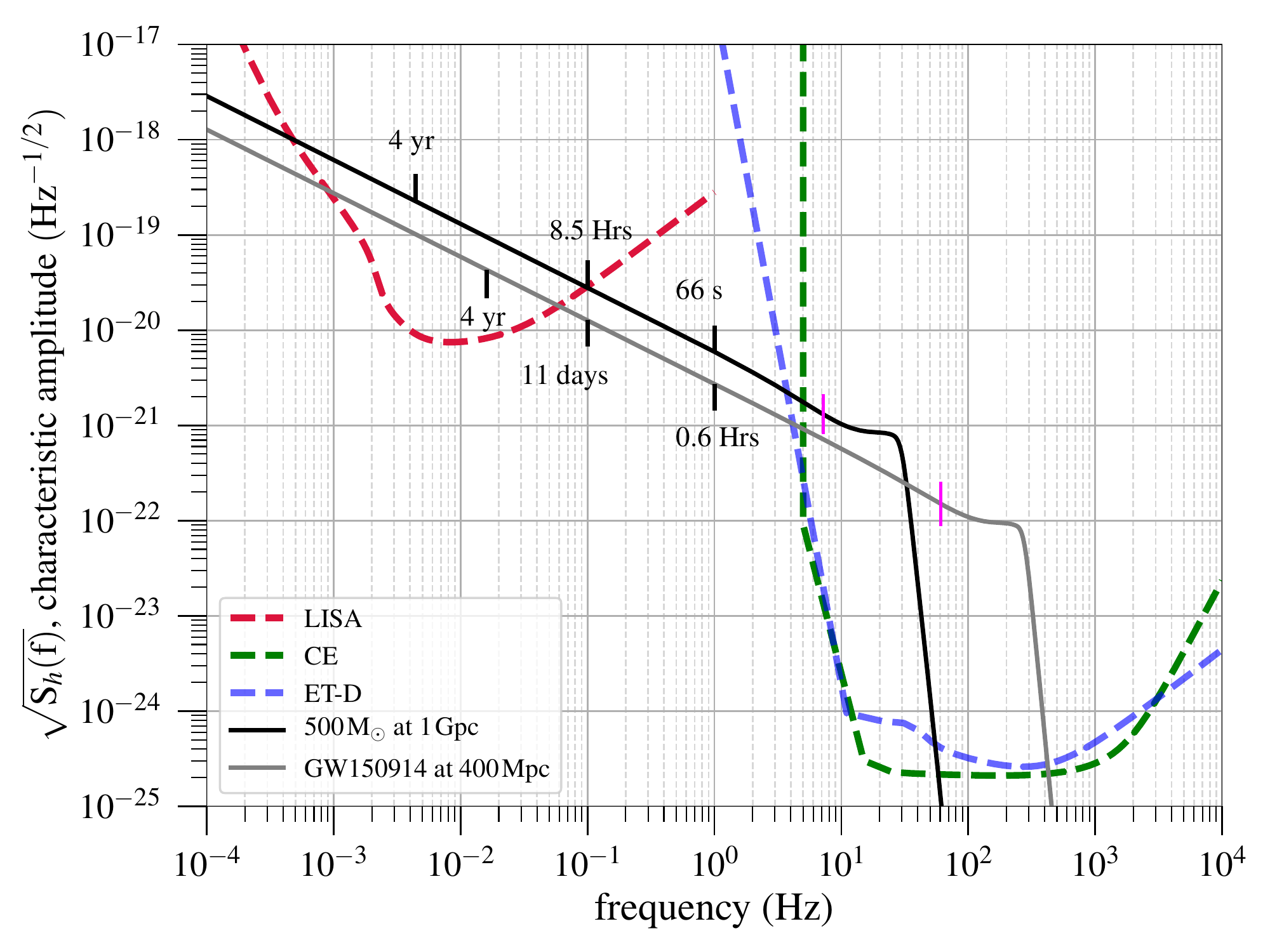}
\caption{The dashed lines denote noise strain sensitivities of LISA, CE and ET. The solid lines denote the characteristic
amplitudes ($2{\sqrt f} \:| \tilde h(f) |$\:) of GW150914-like system at 400 Mpc and an IMBBH system
of $500 M_{\odot}$ at 1 Gpc. The vertical markers in magenta represent the strength of the GW signals from the two sources at the frequency of the last stable orbit. The black markers indicate the time remaining prior to their merger. }
\label{fig:noiseplot}
\end{figure}

Figure~\ref{fig:noiseplot} shows the strain sensitivity (square root of the noise PSD) of LISA, CE and ET. 
It also shows the frequency domain characteristic amplitudes (which signifies the strength of the GW signal)
 of an IMBBH with a total mass of $500 M_{\odot}$ at 1 Gpc and a GW150914-like stellar mass binary black hole system at 400 Mpc for comparison. As the binaries inspiral, the IMBBH spends about a few hours and the stellar mass BBH a few days before leaving the LISA band at 0.1 Hz and entering the ET (CE) range at 3 Hz (5 Hz). The strength of the GW signals at frequencies corresponding to their respective last stable orbits is marked in magenta. 
This shows that the inspiral phase of systems with source-frame masses greater than $500 M_{\odot}$ at a luminosity distance of 1 Gpc, will hardly be visible by CE/ET. The following section discusses this in more detail.
\subsection{Multiband visibility of IMBBH with LISA, CE and ET}\label{sec:mutiband}
\begin{figure}[htp]
\centering
\includegraphics[width=0.5\textwidth]{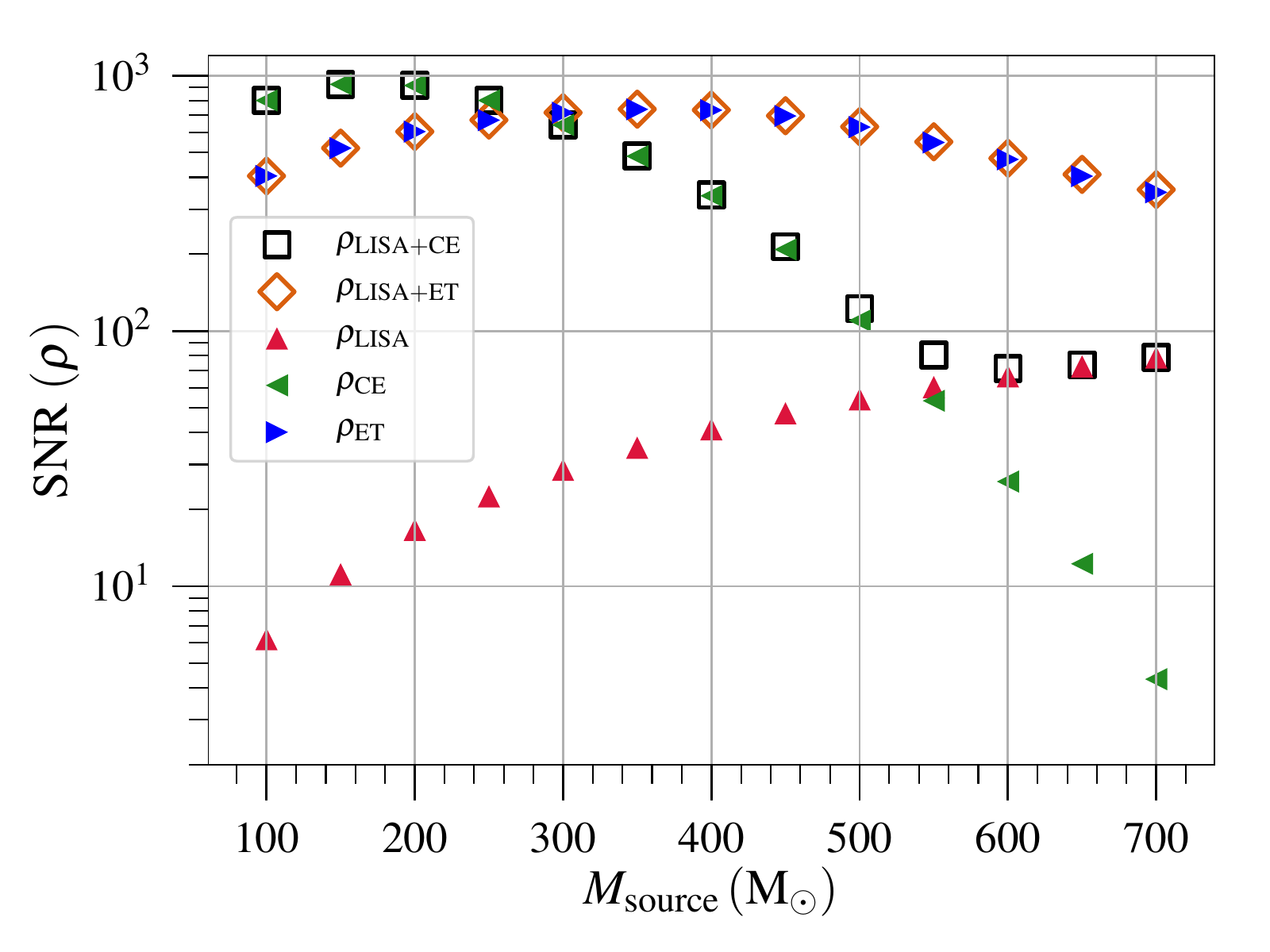}
\caption{SNR from IMBBH sources, accumulated in LISA , CE and ET bands till $f_{\rm LSO}$. All the sources are kept at 1 Gpc with $ q=2,
\chi_1=0.2, \chi_2 =0.1$. As the total mass increases, the visibility of inspiral phase of the IMBBH signal increases
in the LISA band and diminishes in the CE and ET bands. }
\label{fig:snr_lisaceet_flso}
\end{figure}

Following Refs.~\cite{Vitale:2016rfr,Sesana:2016ljz}, there were several works which looked into the detection \cite{Wong:2018uwb,Moore:2019pke} and parameter estimation ~\cite{Grimm:2020ivq,Jani:2019ffg}
 problems in the multiband context. In Ref.~\cite{Jani:2019ffg} the authors
showed that the ET and CE,  which are likely to be operating during the lifetime of LISA  mission, will lead to multiband detections of
IMBBHs up to redshift of ${\sim} 5$. A typical IMBBH with total mass ${\sim}500 M_{\odot}$ at a distance of roughly 1
Gpc will be observable for years in the LISA band when the inspiraling binary components are far apart from each other. This leads to the accumulation of a considerable amount of SNR during the period of observation. 
In Fig.~\ref{fig:snr_lisaceet_flso}, we show the SNR of IMBBHs as a function of total mass for LISA, CE, ET and for multiband observations (LISA+CE and LISA+ET). Though we use IMRPhenomD for the SNR computation,
we integrate the signal up to frequency at the last stable orbit (LSO) corresponding to the total mass $M,$ given by
\begin{equation}
f_{\rm LSO}=\frac{1}{6^{3/2}\pi M } \,.\end{equation}
This choice of the upper cut-off frequency helps in explaining several features in the later sections with regard to the parameterized tests of GR. Note that this choice applies only to the computation of the SNR; for parameter estimation, however, we use the upper cut-off for the full signal as mentioned in Sec.~\ref{sec:CE-ET}.
At a frequency close to 0.1 Hz the IMBBH signal will leave the LISA band and after a few hours, it enters the ET band at around 1Hz and the CE band at 5 Hz as demonstrated in Fig.~\ref{fig:noiseplot}. 
By this time the compact binary will be inspiraling at fairly relativistic speeds until it merges.
The late-inspiral and merger-ringdown phase of the IMBBH evolution will accumulate SNRs of the order 1000 in the CE and ET bands leading to a firm detection.

The SNR in the CE ($\rho_{\rm CE}$) and ET ($\rho_{\rm ET}$) bands initially increases till $200 M_{\odot}$ and $450 M_{\odot}$ respectively and then starts to decrease as the inspiral phase of the system lasts for a shorter and shorter period of time beyond these masses. SNR in the ET band is more than that of the CE from around $350 M_{\odot}$ due to a better low frequency sensitivity of ET between $1-5$ Hz.
Further, the SNR in LISA band ($\rho_{\rm LISA}$) steadily increases and matches with the SNR in CE band roughly at $550
M_{\odot}$. The multiband SNR is defined as $\rho_{\rm MB}=\sqrt{\rho_{\rm
GB}^2+\rho^2_{\rm LISA}}$, where GB denotes a ground-based detector, either CE or ET.

Before we discuss the technical aspects of multiband parameter estimation in the next section, it is important to point out how one
would approach the problem of multibanding in practice. As shown in Fig.~\ref{fig:snr_lisaceet_flso}, IMBBHs would have an SNR of the order of hundreds to thousands in 3G detectors like CE and ET. This would allow estimation of their mass
parameters, especially chirp mass, to incredible accuracies. Using the time of arrival of the signal in the CE band, one would hence be able to search for the low frequency part of the corresponding signal in the LISA band. Due to the prior detection in the ground based detectors, the threshold on SNR for a given false alarm rate could be lowered leading to the detection of sources which have SNR ${\sim} 5$ \cite{Gupta:2020lxa,Ewing:2020brd}. 

A confident detection of an IMBBH in both the LISA and CE detectors forms the basis for multiband parameter estimation. For several of the sources, the IMBBH would have sufficiently high SNR to be visible in the LISA data. For a subset of the sources, an archival search in the LISA band using the information from CE would be required. We find that systems with masses greater than roughly $200 M_{\odot}$ will have sufficiently large SNRs as shown in Fig.~\ref{fig:snr_lisaceet_flso}, hence have the potential of being independently detected by LISA without any requirement of having prior information from CE/ET.

Indeed, more careful studies are required to quantify the detectability of these IMBBHs in the
LISA band (see for instance \cite{Moore:2019pke}), but we do not go into the details of this detection problem in the present work.
\subsection{Multiband parameter estimation}\label{sec:Multiband}
In this section, we discuss various elements of the multiband parameter
estimation we employ in this work.
\subsubsection{Parameter estimation using the Fisher information matrix}
The Fisher matrix is a well known technique used to forecast the statistical uncertainties on various parameters in a parameter estimation
problem when both the signal and noise models are known. When the noise is stationary and Gaussian, in the limit of high SNR, the square
root of the diagonal elements of the inverse of Fisher matrix yields a $1 \sigma$ lower bound on the errors on various parameters~\cite{Rao45,Cramer46,Helstrom68}. In our case, we use this technique to compute the $1 \sigma$ uncertainty
in the estimation of PN deformation parameters that characterize modifications to GR. These errors also translate into upper limits on the values of the deformation parameters for a given detector sensitivity and an increase in the sensitivity would lead
to tighter upper limits. 
As discussed in Sec.~\ref{sec:IMRPD}, our parameter space is spanned by seven GR parameters and up to eight PN deformation parameters, depending on how many deformation parameters are simultaneously estimated in the
problem. The Fisher information matrix is defined as the noise-weighted inner product of the derivatives of the gravitational waveform ${\tilde
h}(f)$ with respect to the parameters ${\theta}^a$ that need to be estimated. More precisely,
\begin{equation}
\Gamma_{mn}=\Big \langle\frac{\partial {\tilde h}(f)}{\partial
\theta^m},\frac{\partial {\tilde h}(f)}{\partial \theta^n} \Big \rangle
\end{equation}
where the noise-weighted inner product is defined as
\begin{equation}
\langle a| b\rangle =4 \,{\rm Re}\int_{f_{\rm low}}^{f_{\rm
\rm high}}\frac{a(f)\,b(f)^{*}+a(f)^{*}\,b(f)}{S_n(f)}\,df,
\end{equation}
where $S_n(f)$ is the noise PSD of the detector and $a(f)$ and $b(f)$ are arbitrary functions of frequency. The lower and upper frequency cut-offs are denoted by $f_{\rm low}$ and $f_{\rm high}$ and depend on the frequency sensitivity bandwidth of the 
detectors, which we discussed in detail in Sec.~\ref{sec:detectors}. 
The Fisher information matrix also allows the use of priors about the parameters provided they are in the form of Gaussian functions~\cite{CF94,PW95}. If $\Gamma^{(0)}$ is the Gaussian prior matrix, the resultant Fisher matrix for any detector is the sum of the prior matrix and the Fisher matrix ($\Gamma_{mn}^{(0)}+\Gamma_{mn}$). The details of prior choices made are discussed in Sec.~\ref{sec:singleparam}.
\subsubsection{Fisher matrix with multibanding for an IMBBH}
For an event jointly detected by a space-based detector (LISA) and a ground-based detector (CE/ET), the multiband Fisher information
matrix is simply the sum of the two Fisher matrices,
\begin{align}
\label{CombFish}
\Gamma_{mn}=\Gamma_{mn}^{\text{GB}}+\Gamma_{mn}^{\text{LISA}},
\end{align}
where $\Gamma_{mn}^{\text{GB}}$ denotes the Fisher matrix corresponding to one of the ground-based detectors, CE or ET. The variance-covariance matrix is defined by the inverse of the multiband Fisher matrix, $$C^{mn}=(\Gamma^{-1})^{mn},$$ where the diagonal components, $C^{mm}$, are 
the variances of $\theta^m.$ The $1 \sigma$ errors on the parameters  $\theta^m$ is, therefore, given as, 
\begin{equation}
\sigma^m = \sqrt{C^{mm}} \,.
\end{equation}
For any IMBBH event observed both in LISA and CE/ET, the errors on each of
the binary parameters returned by the multiband covariance matrix are already marginalized over the rest of the parameters by the very definition of the covariance matrix. However when there is a need to study the variance-covariance matrix of a subspace of the full parameter space (such as the one spanned by the PN deformation parameters that are estimated simultaneously), one can obtain the marginalized matrix by the following well-known prescription of constructing the Schur complement of the Fisher matrix. The Schur complement  of a $p$ dimensional Fisher matrix  $\tilde\Gamma_{p
\times p}$ is given by~\cite{Gallier2019}
\begin{equation}
\tilde\Gamma_{p \times p}= \Gamma_{p\times p} - \Gamma_{p\times q}
\Gamma_{q\times q}^{-1}\left(\Gamma_{p\times q}\right)^T\,,
\label{projection}
\end{equation}
where $\Gamma_{p\times p}$ is the Fisher matrix block corresponding to $\theta_p$ parameters that are of our interest and  for which we want to study the variance-covariance matrix. $\Gamma_{q\times q}$ is the Fisher matrix for $\theta_q$ parameters that we want to marginalize over.
$\Gamma_{p\times q}$ is a matrix with cross terms between $\theta_p$ and $\theta_q$ parameters.
Before we conclude this section, we wish to clarify a subtle issue. As the prior matrix is added to both LISA and CE Fisher matrices, one may suspect over-counting as the prior matrix is featured twice in the multiband Fisher
matrix. But we would like to stress that from the parameter estimation viewpoint, this simply reflects our assumption that parameter estimation with LISA and CE/ET are two independent experiments whose
outcomes are combined to gain greater insights about the dynamics of IMBBHs.This is naturally the way parameter estimation is performed within the framework of Bayesian inference on the data from the two detectors where the posteriors from the parameter estimation of LISA and CE would be combined instead of the likelihoods.
%%%
\section{Effect of multibanding on single-parameter tests of GR}
\label{sec:singleparam}
 In this section, we present the results of our analysis in detail. The first set of results are for the single-parameter tests of GR  where only one of the eight PN deformation parameters is estimated along with
the GR parameters. These are the first estimates of the projected multiband bounds from single-parameter tests of GR with IMBBHs and they complement the earlier works for stellar mass
BBHs~\cite{Carson_2019,Gnocchi_2019}. For the first time, we also provide a concrete explanation, beyond the intuitive arguments,  on why the multibanding improves the tests of GR. 
\subsection{Variation of the bounds with the total mass of the IMBBH:
Single-parameter tests}
\begin{figure*}[t]
\centering
\includegraphics[scale=0.44]{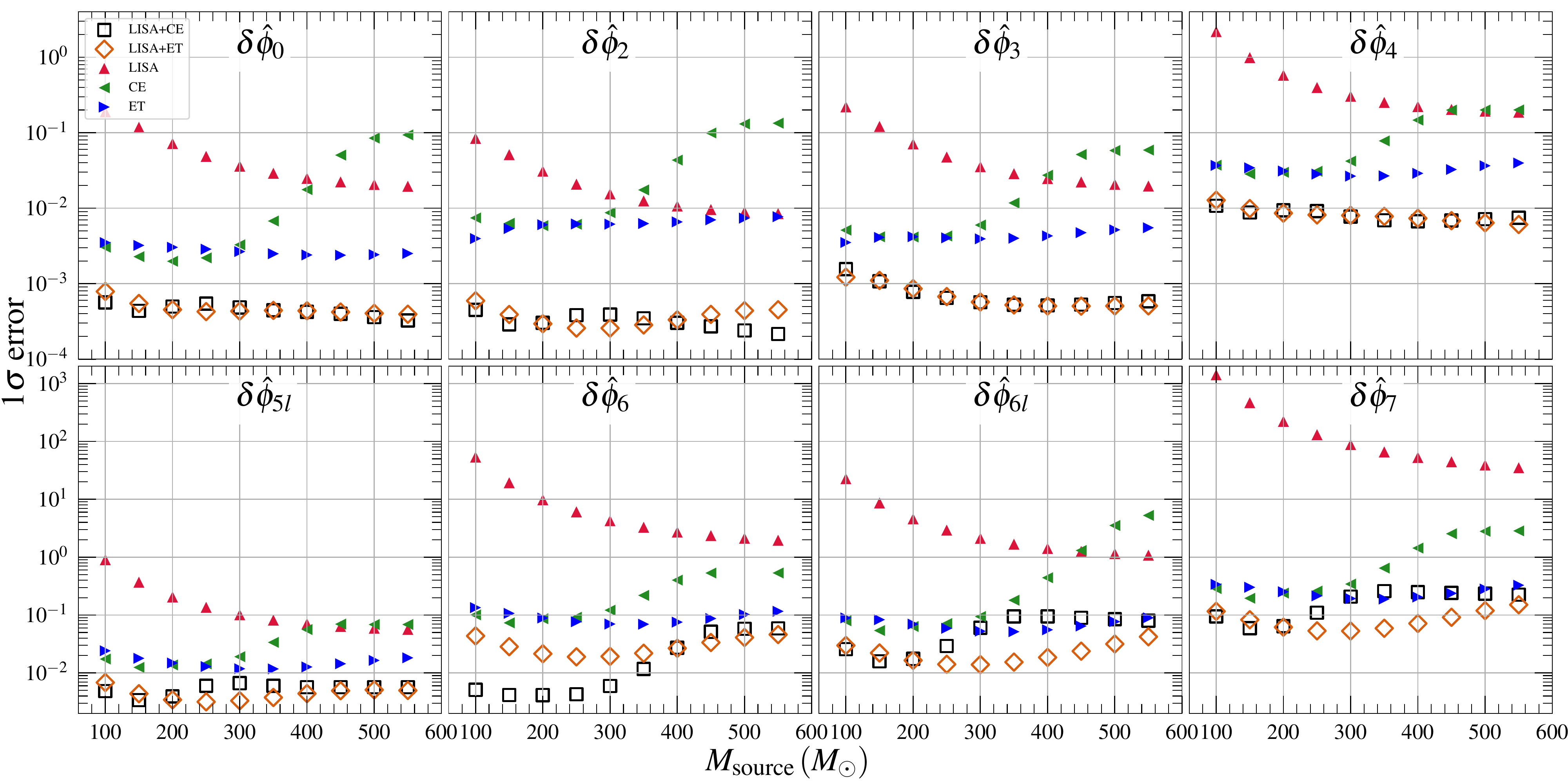}
\caption{Top panel show bounds on deformation parameters at 0PN to 2PN, as a function of total mass in source frame.
Bottom panel show the same but for deformation parameters at 2.5PN to 3.5PN. All the systems have mass-ratio $q=2$, dimensionless
component spins $\chi_1=0.2$ and $\chi_2=0.1$, and luminosity distance $D_L=1\;{\rm Gpc}$.}
\label{fig:oneparam}
\end{figure*}
Our results from single-parameter tests are presented in Fig.~\ref{fig:oneparam} where besides the GR parameters one of the eight PN deformation parameters are estimated at a time. The bounds on the various PN deformation
parameters as a function of the total mass of the IMBBH using only LISA, only CE, only ET and multiband (LISA+CE and LISA+ET) observations are shown in the figure.
For convenience and to facilitate comparison between systems of different masses, the luminosity distance is fixed at 1 Gpc, the mass ratio is 2 and the component spins are $\chi_1=0.2, \chi_2=0.1$.
In our analysis, we employ Gaussian priors on the component spins (mean 0, variance 0.5) and  $\phi_c$ (mean 0, variance $\pi$). The priors on spins and  $\phi_c$ are meant to improve the conditioning of the Fisher
matrices. The maximum source-frame mass for IMBBHs is considered to be $550 M_{\odot}$ because beyond this mass, the signal in the CE band has insufficient inspiral for the parametrized tests to give meaningful results.
This can also be observed from Fig.~\ref{fig:snr_lisaceet_flso} that shows a sharp drop in SNRs carried by the inspiral in the CE band, with increasing total masses of the IMBBHs.

First, let us examine the qualitative features. In Fig.~\ref{fig:oneparam}, the top panel shows  bounds on deformations at the lower PN orders (0PN, 1PN, 1.5PN and 2PN) and the
lower panel depicts the bounds on the higher PN coefficients (2.5PN log, 3PN, 3PN log and 3.5PN). As expected, the lower PN order coefficients are much better constrained than the higher order
ones due to their dominant contribution to the dynamics of the binary. 

The bounds obtained from CE initially decrease with increasing total mass but start to increase slowly after ${\sim} 200 M_{\odot}$ following an inverse of the SNR trend expected from the Fisher matrix. 
The bounds obtained with ET stay comparable throughout the mass range with the existence of a weak minimum around the loudest system in  ET band which is $\sim 350 M_{\odot}$ as shown in Fig.~\ref{fig:snr_lisaceet_flso}.
The errors from LISA almost monotonically decrease with increasing mass following the inverse of the SNR trend. Furthermore, one also observes cross-overs between the LISA, CE, and/or ET curves for most of the deformation parameters, which is simply an imprint of the similar cross-overs seen in the SNR curves in Fig.~\ref{fig:snr_lisaceet_flso}. The total masses at which these cross-overs happen are different for the SNR
and the deformation parameters as the latter have more complicated noise moments~\cite{PW95} (powers of frequency weighted by the noise PSD) that constitute the Fisher matrix.  
The multiband bounds with LISA+ET are comparable to that of LISA+CE, especially for the lower PN deformation parameters. Some improvements are observed in the multiband bounds on the two highest PN orders (3PN log and 3.5PN) when ET is used instead of CE.

We notice that bounds on $ \hat{\delta\phi_0}, \hat{\delta\phi_2}$ and $\hat{\delta\phi_3}$ improve the most due to multiband observations for the entire IMBBH total mass range considered. LISA alone helps in constraining lower PN orders more than the higher PN orders as it collects more information in the low frequency regime. However, CE and ET have more information than LISA on the high frequency PN orders and can constrain all of them to $\leq {\cal O}(1$) accuracy. The bounds on $ \delta\phi_0, \delta\phi_2$ and $\delta\phi_3$ improve approximately by a factor of 20 to 70 (10 to 30) for masses greater than $300 M_{\odot}$, upon combining Fisher matrices for LISA and CE (ET).
However, the multiband bounds on higher PN deformation parameters, $\delta\phi_{4}$, $\delta\phi_{5l}$, $\delta\phi_{6}$, $\delta\phi_{6l}$ and $\delta\phi_{7}$ mostly follow CE/ET except at higher masses where multiband observations improve the bounds by a factor of 5 to 10 due to the high SNR in the LISA band. These substantial improvements in bounds, specifically at the higher mass regimes, may come as a bit of surprise as the multiband bounds are factor of tens better than the best bounds obtained from either of the detectors. We devote the next subsection to explaining this interesting result.

We also find that we get the best multiband bounds on PN deformation parameters for the equal mass case, $q=1$ and the bounds worsen with increasing mass asymmetry, though not drastically. The multiband bounds also improve with increasing dimensionless spin magnitudes, particularly for higher PN orders. 

We conclude this subsection with some quantitative statements that can be read from Fig.~\ref{fig:oneparam}. 
The best bounds due to multibanding are obtained on 0PN and 1PN phase deformation parameters that are roughly measured to $\leq {\cal O}(10^{-3})$ accuracy. 
 The rest of the parameters can be estimated to roughly between ${\cal O}(10^{-3})$ and ${\cal O}(10^{-1})$ accuracy for IMBBHs of total mass ranging from $100-550 M_{\odot}$ at 1 Gpc.
 \subsection{Explaining the improvement due to
multibanding}\label{sec:explainMB}

As we next discuss the effect of multibanding, we consider only CE as a representative of 3G ground-based detectors. This is because the differences between ET and CE are small enough and it would not make any difference to our conclusions. Figure~\ref{fig:oneparam} shows that the multiband observations can provide huge  improvements in the bounds of $\delta\hat{\phi}_{0}$, $\delta\hat{\phi}_{2}$, and $\delta\hat{\phi}_{3}$ in spite of the low SNRs in LISA band. For instance, bounds on $\delta{\hat \phi}_0$ from LISA and CE at $500 M_{\odot}$ are ${\cal O}(10^{-2})$ as compared to the joint multiband bound which is ${\cal O}(10^{-4})$. This 2-order-of-magnitude improvement may seem surprising at first. However, our investigations reveal that this feature is due to the cancellation of several off-diagonal terms (which correspond to degeneracies in the parameter space) when we add the Fisher matrices of LISA and CE to obtain the multiband Fisher. Due to this cancellation, the inverse of this combined Fisher matrix results in errors that are significantly smaller than the ones from LISA or CE alone. 

Owing to the difficulties in representing higher dimensional matrices pictorially, we focus on selected two-dimensional subspaces, which are highly correlated and the corresponding ellipses to understand the effect of multibanding. 
Consider  an IMBBH system of total mass $500\ M_{\odot}$ with spins $\chi_1=0.2, \chi_2=0.1$ at a distance of 1 Gpc. It is intuitive to relate the area of the two dimensional ellipses for a particular detector to its ability to
simultaneously measure the two parameters. The smaller the area, the better is the measurement. Likewise, the orientation or the tilt of the ellipses tells us about the sign of the correlation between the two
parameters: positively tilted ellipses (whose semi-major axis subtends an angle less than 90 degrees) refer to a positive correlation between the two parameters and negatively titled ellipses indicate
a negative correlation between the two parameters. 

\begin{figure}[h]
\centering
 \includegraphics[width=1\linewidth]{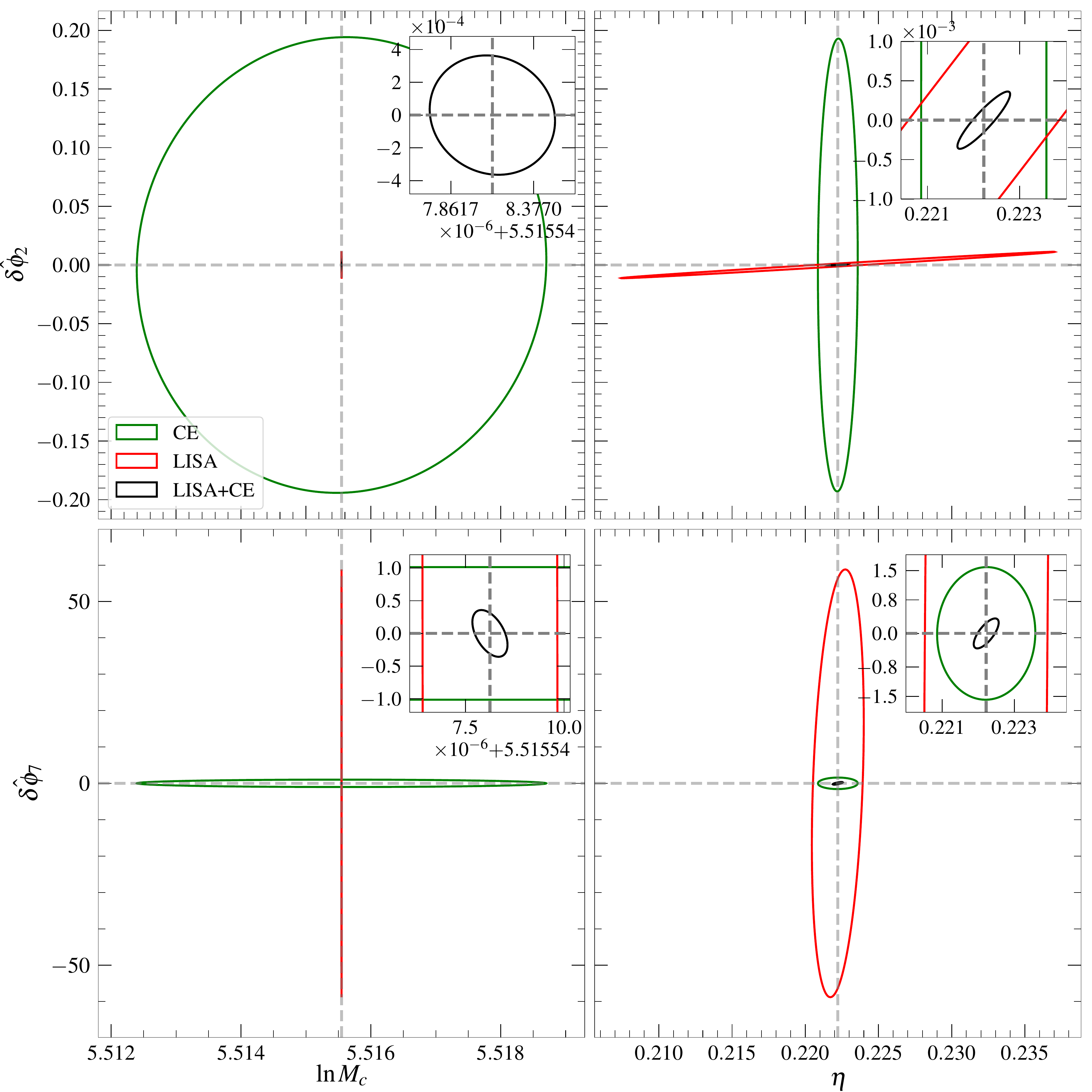}
\caption{The top panels consist of two-dimensional  contour plots between ${\rm ln} M_c-\delta{\hat \phi}_2$ and $\eta-\delta{\hat \phi}_2$. The bottom panels show two-dimensional contour plots between ${\rm ln} M_c-\delta{\hat \phi}_7$ and $\eta-\delta{\hat \phi}_7$. The red and green contours correspond to LISA and CE respectively. The multiband contours that fall inside the intersection of LISA and CE contours are all shown in the insets of their respective plots in black.}
\label{fig:contour}
\end{figure}

Let us consider  $\delta \hat{\phi}_2$  and $\delta\hat{\phi_7}$ estimates for the demonstration, as multiband improvement is the highest for the former and the lowest  for the latter. Figure~\ref{fig:contour}  shows the
$1\sigma$-confidence ellipses of $\delta \hat{\phi}_2$ (top panel) and $\delta\hat{\phi}_7$ (bottom panel) with $\ln M_c$ (left panel) and $\eta$ (right panel). These two-dimensional ellipses are obtained by marginalising over the
remaining parameters following the prescription in Eq.~(\ref{projection}).
The error ellipsoid for CE in the  $\ln M_c - {\delta {\hat \phi}_2}$ plane has an area much larger than that of LISA and a tilt that is orthogonal to the one for LISA.
This implies that LISA can measure both the parameters better than CE and the correlation between these two parameters is negative for LISA and positive for CE. The multiband two-dimensional confidence ellipse is in black, whose
zoomed-in version is shown in the inset. The area of this ellipse is substantially smaller than that of the LISA and CE. Table~\ref{tab:area} provides the areas of the two-dimensional ellipses
for various parameter combinations. This is a quantitative demonstration of the complementarity of the two frequency bands very often invoked in the literature to explain the improvements due to multibanding \cite{Cutler:2019krq}. 
From the confidence ellipses shown in the $\eta - {\delta {\hat \phi}_2}$ plane, we can say that LISA seems to be able to measure ${\delta {\hat \phi}_2}$ better than CE. On the other hand, CE's ability to measure $\eta$ is better than LISA.

These intriguing features may have a more intuitive explanation in terms of the structure of the phasing formula.
The low frequency sensitivity of LISA helps to measure the lower order PN parameters. 
For instance, LISA measures chirp mass, $M_c$ to very high accuracy, because of which the correlation between $\delta\hat\phi_2$ and  $M_c$ is weaker.
Hence, LISA can measure both $M_c$ and  $\delta\hat\phi_2$ with accuracies better than CE.
Combining information from both these detectors facilitates further weaker correlations between $M_c$ and  $\delta\hat\phi_2$ and improves measurement on both these parameters. 
As mentioned earlier, this improvement is due to the cancellation of off-diagonal terms of the Fisher matrix owing to complementary signs of correlations. 
Since the marginalization also involves cross-terms in the covariances matrix, the two-dimensional plots are sensitive to such cancellations.
Now, LISA is not very efficient in breaking the degeneracy between $\delta\hat\phi_2$ and $\eta$. 
We know that $\eta$ first appears at 1PN order in the phasing formula and $\delta\hat\phi_2$ parametrizes deviation at the same order. 
In order to break this degeneracy, sensitivity to higher order PN phasing coefficients is necessary which LISA does not have.
 On the other hand, CE does have this capability and is hence able to break this correlation better.  
Hence, LISA estimates $\delta\hat\phi_2$ better than CE ($\delta\hat\phi_2$ being a low frequency term) but it is CE that estimates $\eta$ better than LISA. 
Once again, this complementarity of LISA and CE leads to an overall improvement of bounds on $\delta\hat\phi_2$ and $\eta$ on multibanding. 

Similar two dimensional ellipses are presented for   $\ln M_c - \delta\hat{\phi_7}$ and $\eta - \delta\hat{\phi_7}$ in the bottom panels of 
Fig.~\ref{fig:contour}. Here, as one may read off from
Table~\ref{tab:area}, the $\ln M_c - \delta\hat{\phi_7}$ ellipses have the smallest area for LISA, though the two areas 
do not differ as much as they did for $\ln M_c - \delta\hat{\phi_2}$ case. 
As $\delta\hat\phi_7$ is a deformation of a higher PN coefficient, 
it is measured very poorly with LISA whereas $M_c$ is very well measured as we saw earlier. 
On the other hand CE measures both $\delta\hat\phi_7$ and $M_c$ very well though $M_c$ measurement is not as good as LISA, as expected. 
Due to the better sensitivity to higher PN orders, CE is also able to break the correlation between $\delta\hat\phi_7$ and $\eta$ better so as to measure both the parameters to good accuracies. 
The joint LISA + CE bounds, again, immensely benefit from these two individual measurements and help to get rid of correlations, thereby leading to significantly improved joint bounds. 

 To summarize, we have explicitly shown that the huge improvements due to multibanding is a direct consequence of the cancellation of various correlations between parameters when we combine the information from LISA and CE. 
As LISA is sensitive to the early inspiral phase of the evolution of IMBBHs, it is most sensitive to the lower-PN-order deformation parameters. Cosmic Explorer, on the other hand, is sensitive to the late inspiral and merger-ringdown phases and hence sensitive to the higher-PN-order deformation-parameters. 
Both the low- and high-PN coefficients are strongly correlated with the mass parameters. However, the signs of the correlations depend on the frequency of observation and hence can be opposite for LISA (which is sensitive to frequencies less than 1 Hz) and CE (which is sensitive to frequencies above 1 Hz).  The addition of the Fisher information matrices from the two independent measurements can hence cancel the off-diagonal elements due to their opposite signs, leading to a resultant Fisher matrix, whose elements are much less correlated compared to the individual ones. As this happens with several of the cross-terms of the Fisher matrix, with varying degrees of efficiency, the resultant multiband bounds are substantially improved.  A simplistic example for understanding this is given in the Appendix.
\begin{table}
\begin{tabular}{|c|c|c|c|}
\hline
Parameters & LISA & CE & LISA+CE\\
\hline
$\; {\rm ln} M_c-\delta{\hat \phi}_2\; $  & \;\; $ 8.5 \times 10^{-8}$ \;\;&\;\; $1.9 \times 10^{-3}$ \;\;&\;\; $4.5 \times 10^{-10}$\;\;\\ \hline
$\; \eta-\delta{\hat \phi}_2\; $  & \;\;$4.8 \times 10^{-5}$ \;\;&\;\; $8.2 \times 10^{-4}$\;\; &\;\; $2.8 \times 10^{-7}$\;\;\\ \hline
$\; {\rm ln} M_c-\delta{\hat \phi}_7 \;$  &\;\; $3.3 \times 10^{-4}$\;\; &\;\; $0.01$\;\; &\;\; $4.4 \times 10^{-7}$\;\;\\ \hline
$\; \eta-\delta{\hat \phi}_7\; $  &\;\; $0.33$\;\; &\;\; $6.7 \times 10^{-3}$\;\; &\;\; $2.7 \times 10^{-4}$\;\;\\
\hline
\end{tabular}
\caption{Area of the $1\sigma$ confidence ellipses shown in Fig.~\ref{fig:contour}.}
\label{tab:area}
\end{table}
%%%
\section{Multiparameter tests of GR with IMBBH}
\label{sec:multiparam}
So far, we have presented the bounds on the PN deformation parameters when only one of them is estimated at a time. 
There are eight PN coefficients up to 3.5PN order which give us eight single-parameter tests of GR. As argued earlier, the most general
test of GR we can carry out is the one where all of the PN deformation parameters are {\it simultaneously} measured. Due to the
inherent degeneracies between the deformation parameters and the GR
parameters, this test is not feasible using GW observations in  a single frequency band as argued
in \cite{AIQS06a}. 

We now study how a set of  PN deformation parameters may be measured simultaneously by combining observation of the same signal in LISA and CE
bands. We consider two different types of multiparameter tests of GR. In the first type, we increase the number of parameters that occur at
different PN orders starting from the lowest (0PN) order. The second type of tests measure the deviation in the PN coefficients starting from the
highest (3.5PN) order and going to the lower PN orders. Hence this set of tests would quantify our ability to
constrain possible deviations for those theories which predict the {\it last} $n$ coefficients to differ from GR. For instance, $n=3$ would be a test where we simultaneously measure the
last three PN parameters in the phasing formula. Below we provide the projected bounds on a prototypical IMBBH
system with a total redshifted-mass of $200 M_{\odot}$ at a luminosity distance of 1 Gpc for these two subclasses of
multiparameter tests.

From Fig.~\ref{fig:oneparam} it is evident that in the case of single-parameter tests, the total source frame
mass range in which the joint bounds on most of the parameters are minimum is around $\sim150-300 M_{\odot}$.
This can be understood from two features which have already been discussed: (i) the trends in the SNR as
a function of total mass, and (ii) higher PN coefficients play a dominant role in the late time dynamics, close to the merger, which
falls in the CE band. The effectiveness of a multiparameter test of GR, or for that matter any test of GR, depends on the optimization 
of these two effects, which leads to a sweet spot for these tests. In our case, this happens to be at a total mass of $\sim163 M_{\odot}$ 
(total redshifted-mass of $200 M_{\odot}$ at 1Gpc). The errors on different deformation parameters in
Fig.~\ref{fig:oneparam} show slightly different minima, hence there are other values around $200 M_\odot$
that are equally good for these tests.  As this choice would have a negligible impact on the conclusions we draw, we
stick to a total redshifted-mass of $200\,M_{\odot}$ with two different choices of component spins to show the
projected bounds using multiparameter tests. 
\subsection{Bounds from the lower-order PN side}
\begin{figure}[htp]
\centering
\includegraphics[width=0.49\textwidth]{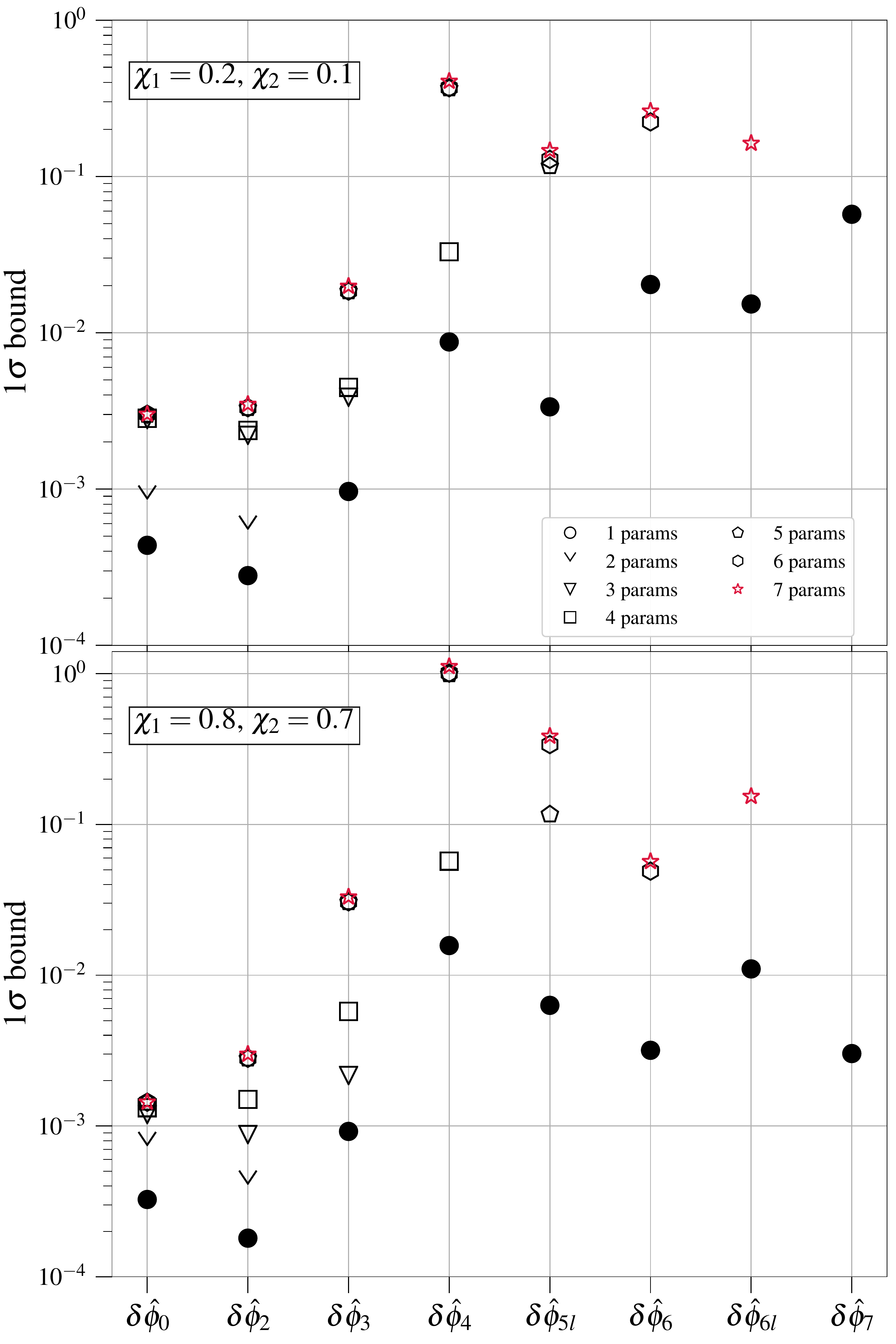}
\caption{Bounds on PN deformation parameters from $n$-parameter tests starting from 
         0PN through 3.5PN with an IMBBH system of total redshifted-mass of $200 M_{\odot}$ at 1 Gpc, with mass ratio 2
        and two different spin configurations, $\chi_1=0.2, \chi_2=0.1$ (upper panel) and $\chi_1=0.8, \chi_2=0.7$ (lower panel).}
\label{fig:goldenbinary}
\end{figure}

Figure~\ref{fig:goldenbinary} shows the bounds on deformation parameters
obtained from the various $n$-parameter tests starting from 0PN, for an IMBBH with a total 
redshifted-mass of $200 M_{\odot}$ and a mass ratio of 2 with two different spin
configurations: $\chi_1=0.2,\chi_1=0.1$ (top panel) and $\chi_1=0.8, \chi_2=0.7$ (bottom panel).
As there are seven GR parameters, for each test, we invert a Fisher matrix of dimension $7+n$ to obtain
the corresponding errors. We find that simultaneous measurement of only
seven of the eight PN deformation parameters is possible for this binary
configuration if we require the errors on all the PN deformation parameters 
to be less than or equal to unity.  Comparing the top and the bottom panels,
it is evident that the increase in spin magnitudes have varied effects on the 
estimation of PN coefficients, sometimes improving and other times worsening the error bounds.
The lower order PN coefficients are also largely unaffected by spin magnitudes. This is a reflection
of the fact that spin effects are higher order effects (starting at
1.5PN order) and hence spin dynamics plays a dominant role in the late time
dynamics where, again, CE sensitivity has an important role, leading to improved bounds
of the higher order PN deformation parameters.

For high values of spin magnitudes, single-parameter tests on the Newtonian and 1PN
coefficients would yield constrains of ${\cal O}(10^{-4})$ while
all other parameters, except the 2PN and 3PN logarithmic ones, will be
bounded to ${\cal O}(10^{-3})$. The worst bounds are for the 2PN and 3PN
logarithmic terms which are of ${\cal O}(10^{-2})$. This precision is
unprecedented compared to what LISA and CE would be able to do for
supermassive and stellar mass BBHs, respectively, to which they are most
sensitive to \cite{Gupta:2020lxa}. By increasing the
number of parameters that are simultaneously measured, the bounds on all
the lower order PN deformation parameters worsen due to the degeneracies
present in the waveforms. The addition of the 
$\delta {\hat \phi}_{\rm 5l}$, $\delta {\hat \phi}_{\rm 6}$, and $\delta {\hat \phi}_{\rm 6l}$ deformation parameters
have negligible effects on the bounds on
the lower order PN deformation parameters as the correlations of these lower order parameters with
the higher order ones are rather weak. However, adding the 3.5PN parameter
significantly worsens the bounds on all the deformation parameters above 2PN, making
the errors go above unity and hence these are not shown. This trend is a consequence of
the superior ability of LISA to measure the lower order PN coefficients and CE to
measure the higher order ones. As we keep adding higher order PN coefficients,
the bounds on the lower order ones, which benefit mostly from LISA, are
unaffected. However, when we add more higher order PN parameters, such as
3.5PN, CE's ability to simultaneously measure them diminishes, 
leading to an overall worsening of higher order PN deformation parameters' measurement. Despite this,
it is impressive to note that, with multibanding, a seven parameter test of GR can yield bounds of
$\leq {\cal O}(1)$ for all seven PN deformation parameters. 
\begin{figure}[htp]
\centering
 \includegraphics[width=0.49\textwidth]{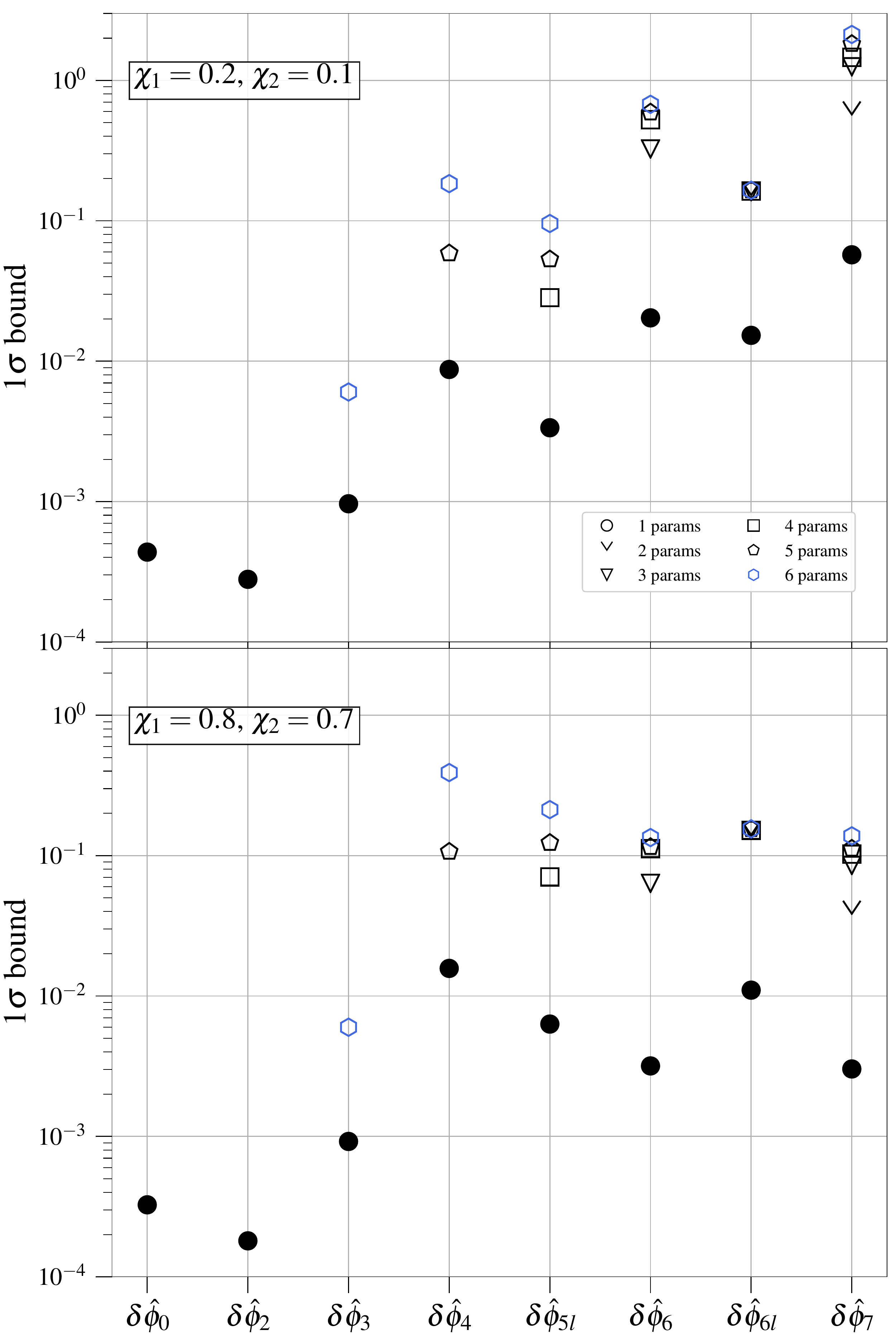}
\caption{Bounds on PN deformation parameters from $n$-parameter tests starting 
         from 3.5PN through 0PN with an IMBBH system of total redshifted-mass of $200 M_{\odot}$ at 1 Gpc, with mass ratio 2 and two different spin configurations, $\chi_1=0.2, \chi_2=0.1$ (upper panel) and $\chi_1=0.8, \chi_2=0.7$ (lower panel).}
\label{fig:goldenbinaryright}
\end{figure}
\subsection{Bounds from the higher-order PN side}
Figure~\ref{fig:goldenbinaryright} shows the second set of results related
to the bounds on the highest $n$ PN deformation parameters. For instance, one can constrain any
modified theory of gravity which predicts deviations from GR starting
at 2PN order (five parameter bounds denoted by pentagons) with a
precision ${\cal O}(10^{-1})$. 

Though CE is mostly sensitive to the higher order PN coefficients, it is not capable of 
breaking the degeneracy between two consecutive PN coefficients
which are strongly degenerate. In this case, LISA is also able to offer
little help as its role is limited to estimating $M_c$ and $\eta$ very well 
and breaking their degeneracies with higher PN coefficients. As these correlations are rather
weak, LISA does not help much for this class of tests. 
%%%
\subsection{Implications of GW190521}
As noted earlier, the large redshift of GW190521 makes it unlikely for this event to be a multiband candidate.  Hence we consider a system with the same source-frame-mass as 
GW190521, but at 1 Gpc, for a case study.  This would mean that the redshifted component masses have to be $\sim 104 M_{\odot}$ and $\sim 80 M_{\odot}$.  
For such a binary, SNRs in LISA and CE are $\sim12$ and $\sim1565$, respectively. This closely resembles the case that we considered in Sec.~\ref{sec:multiparam} and the projected multiband multiparameter bounds are similar.
\section{Caveats}\label{sec:caveats}
In this section we discuss some of the caveats of our analysis.

{\bf Neglect of modifications to certain PN-order terms in the phasing formula:} We have considered only the modifications of those phasing coefficients which are nonzero in GR. For instance, there are certain theories of gravity which predict modifications to (effective) negative PN orders (see Table 1 of \cite{Chamberlain:2017fjl}) which are not considered here. Further, considering them would significantly enlarge the parameter space. In the same spirit, we have also not considered any deviation at 0.5PN order in the phasing, which again is absent in GR. 

{\bf Uncertainties about the IMBBH population:} 
An important caveat of the results presented here is the uncertainty in the merger rates of IMBBHs. 
The discovery of GW190521 has led to the rate of similar systems to be  $0.13^{+0.3}_{-0.1} \,{\rm Gpc}^{-3}\,{\rm yr}^{-1}$. This implies that in the LISA-3G era, there will be a few events like GW190521 but nearer (say, at 1 Gpc).
The upper limits from the LIGO/Virgo searches for IMBBHs are reported in Ref.~\cite{Salemi:2019ovz}. 
Hence an estimate about the masses of IMBBHs for which this test would perform well will be clearer only in the future 
with more detections of high mass BBHs and/or more stringent upper limits.

{\bf Use of Fisher-matrix-based parameter estimation:} We have relied on the ability of the
Fisher information matrix approach to
predict the precision with which the PN deformation parameters can be
estimated using LISA and CE/ET. The Fisher-matrix-based approach is expected
to be reliable in the limit of high
SNR~\cite{CF94,Balasubramanian:1995bm,Vallisneri07}.
However, the projected bounds on non-GR parameters from the Fisher matrices
for GW150914 and GW151226 have shown good agreements with the results
from Bayesian inference~\cite{YYP2016} reinforcing the utility of the Fisher
matrix to obtain an order of magnitude estimates of the errors. In our case, the SNRs in the CE
band are of the order of hundreds to thousands and hence are well within
the domain of applicability of the Fisher matrix. However, the LISA SNRs are of
order $\geq10$ which theoretically falls only marginally within the
domain of applicability of this method.  Hence our LISA-only results are
prone to have uncertainties which need to be quantified
using numerical sampling techniques such as Markov chain Monte Carlo \cite{Metropolis} or Nested Sampling \cite{Skilling}. 
 A recent work \cite{Toubiana:2020vtf} has paved the way for more work in this direction.

{\bf Neglect of precession, eccentricity and subdominant modes in the gravitational waveforms:} The bounds reported in this paper were obtained using the IMRPhenomD waveform model, which models a non-precessing black hole binary inspiraling in quasi-circular orbits. This model does not account for effects such as spin-induced orbital precession~\cite{ACST94} or subdominant modes of the GW signal~\cite{BIWW96,ABIQ04,ABFO08,MKAF16}. The incorporation of precession~\cite{Khan:2018fmp} and subdominant modes~\cite{Khan:2019kot} bring in characteristic modulations to the phase and amplitude of the waveform and hence is believed to be more informative in improving the overall parameter estimation (see, for instance, \cite{ChrisAnand06b,AISSV07,MAIS10,SW09,HKJ11}). Therefore, one would expect our bounds to improve with the incorporation of these effects, however, this is outside the scope of the paper. Moreover, IMRPhenomD does not account for eccentricity either. This is a drawback of our analysis as binaries during their early inspiral phase when in the LISA band, could be on eccentric orbits.

{\bf Neglect of LISA's orbital motion}: Our model for the response function for LISA does not account for its orbital motion. As these orbital modulations have negligible impacts on the estimation of intrinsic parameters of the binary~\cite{BBW05a,ALISA06}, our estimates are unlikely to be affected much by this assumption.
%%%
\section{Conclusions}\label{sec:conclusion}
Future ground-based and space-based detectors would detect several intermediate mass BBHs and a subset of them would be visible in both bands~\cite{Jani:2019ffg,Bellovary:2019nib,Cutler:2019krq}. Such multiband detections can have important implications for tests of GR.
 We have discussed in detail the possibility of multiband observation of IMBBH systems using 3G ground-based detectors
and the space-based detector LISA. It is shown that observations of IMBBHs would be an excellent new class of sources for tests of the 
strong-field dynamics. Besides the single-parameter tests of GR, IMBBHs would facilitate multiparameter tests, which simultaneously measure more than one PN deformation parameter.

The addition of information from LISA, which is sensitive to the lower PN orders, and CE/ET, which is sensitive to higher PN orders, 
leads to significant improvements in the bounds on deformation paramters. This is due to the massive cancellations of the 
off-diagonal entries of the Fisher matrix which signify how multibanding helps to break the degeneracies between 
various parameters in the gravitational waveform. 
We have discussed how the projected bounds would vary as a function of the total mass of the system and 
find that an IMBBH with a total redshifted-mass of $200M_{\odot}$ would be on the sweet spot for multiband multiparameter tests.  
This system of mass ratio 2  at 1 Gpc, can measure all the PN coefficients for single-parameter tests to below 10\% and can simultaneously 
estimate first seven PN coefficients to below 50\%.
%%%
\section*{Acknowledgments} 
We thank Emanuele Berti, Ssohrab Borhanian, Arnab Dhani, Muhammed Saleem, and especially Ajit Mehta for useful discussions and comments.  B.S.S. is supported in part by NSF Grants No. PHY-1836779, No. AST-1716394, and No. AST-1708146. S.D. and K.G.A.  also acknowledge partial support by the Swarnajayanti Grant No. DST/SJF/PSA-01/2017-18 of DST-India.  K.G.A. also acknowledges partial support by SERB Grant No. EMR/2016/005594 and a grant from the Infosys Foundation. This document has LIGO preprint No. {\tt P2000209}. 

\appendix
\section{A toy model for understanding the improvement due to Multibanding }\label{appendix}
To demonstrate how multiparameter tests are facilitated via breaking of degeneracies between parameters due to multibanding consider two $2\times2$ matrices, $\Gamma_{LISA}$ and $\Gamma_{GB}:$ 
\begin{equation}
 \Gamma_{LISA}= \begin{bmatrix}
\Gamma_{xx}^{\rm L} &\Gamma_{xy}^{\rm L}\\
\Gamma_{xy}^{\rm L} & \Gamma_{yy}^{\rm L} 
\end{bmatrix}\\
, \Gamma_{GB}= \begin{bmatrix}
\Gamma_{xx}^{\rm G} &  \Gamma_{xy}^{\rm G}\\
 \Gamma_{xy}^{\rm G}& \Gamma_{yy}^{\rm G}
\end{bmatrix}
\end{equation}
These can be thought as 2D Fisher matrices for two parameters obtained with LISA and with one of the 3G detectors, respectively.  The multiband Fisher matrix, $\Gamma_{MB}$ can be obtained by adding the two, 
\begin{equation}
\Gamma_{MB}= \Gamma_{LISA}+\Gamma_{GB}.
\end{equation}
Inverting the multiband Fisher matrix $\Gamma_{MB}$ gives the multiband variance-covariance matrix $C_{MB}=\Gamma_{MB}^{-1}$.  The multiband bounds are the square roots of the diagonal elements of $C_{MB}$. Straightforward algebra gives the errors on the two variables as 
\begin{equation}
\sigma_{x}^{MB}= \sqrt{\frac{1}{(\Gamma_{xx}^{\rm L}  +\Gamma_{xx}^{\rm G} )-(\Gamma_{xy}^{\rm L} +  \Gamma_{xy}^{\rm G})^2}},
\label{sigmaMBx}
\end{equation}
\begin{equation}
\sigma_{y}^{MB}= \sqrt{\frac{1}{( \Gamma_{yy}^{\rm L}+\Gamma_{yy}^{\rm G})-(\Gamma_{xy}^{\rm L} +\Gamma_{xy}^{\rm G} )^2}}.
\label{sigmaMBy}
\end{equation}

Equations (\ref{sigmaMBx}) and (\ref{sigmaMBy}) clearly show that opposite signs of the cross-terms $ \Gamma_{xy}^{\rm L}$ and $ \Gamma_{xy}^{\rm G}$ could make the joint bounds much 
smaller than the bounds obtained from individual detectors. Due to the differences in the signs of many of the Fisher elements of LISA and GB, cancellations of the cross-terms lead to a dramatic improvement in the measurement of non-GR parameters as shown in Sec.~\ref{sec:multiparam}.

\bibliographystyle{apsrev}
\bibliography{../Ref_list}

\end{document}